\documentclass[aps,prl,onecolumn,groupedaddress,superscriptaddress,showpacs]{revtex4}
\usepackage{amssymb}
\usepackage{amsmath}
\usepackage{graphicx}
\usepackage{epstopdf}

\begin{document}

\title{Terahertz Electrodynamics of $180^\circ$ Domain Walls in Thin
Ferroelectric Films}
\author{Igor Lukyanchuk}
\affiliation{University of Picardie, Laboratory of Condensed Matter Physics, Amiens,
80039, France}
\affiliation{Materials Science Division, Argonne National Laboratory, Argonne, Illinois
60439, USA}
\author{Alexey Pakhomov}
\affiliation{University of Picardie, Laboratory of Condensed Matter Physics, Amiens,
80039, France}
\affiliation{Department of Experimental Physics, Voronezh State University - Voronezh
394000, Russia}
\author{Ana\"{\i}s Sen\'{e}}
\affiliation{University of Picardie, Laboratory of Condensed Matter Physics, Amiens,
	80039, France}
\author{Alexandr Sidorkin}
\affiliation{Department of Experimental Physics, Voronezh State University - Voronezh
394000, Russia}
\author{Valerii Vinokur}
\affiliation{Materials Science Division, Argonne National Laboratory, Argonne, Illinois
60439, USA}

\begin{abstract}
We investigate oscillation dynamics of a periodic structure of the $%
180^\circ $ domain walls in nanometricaly thin substrate-deposited
ferroelectric films and superlattices. We calculate dynamic permittivity of
such structures and reveal a collective resonance mode, which in the typical
ferroelectric compounds, PbTiO$_3$/SrTiO$_3$, lies in the sub- and low THz
frequency range of $0.3 \div 3\,\mathrm{THz}$. We propose the
reflection-absorbtion spectroscopy experiments to observe this mode.
\end{abstract}

\pacs{77.80.Dj, 62.25.Fg, 62.25.Jk, 77.22.Ch, 77.55.fg}
\maketitle


A high-frequency dynamic response of ferroelectric materials is widely used
in advanced electronic and optoelectronic applications. A wide range of
operational frequencies that spans several decades from few kilohertz to
tens terahertz ($1\,\mathrm{THz}$ $=10^{12}\,\mathrm{Hz}$) is provided by
the multiscale organization of spontaneous polarization. Whereas the ac
response properties in radio- and microwave diapasons $\ 10^{4}\,\mathrm{Hz}%
\div 0.3\,\mathrm{THz}$ are mostly due to the relaxation dynamics of the
domain walls (DWs) and polar clusters, the far-infrared spectral region $\ $%
of $3\div 30\,\mathrm{THz}$ is governed by the soft-mode vibrations of the
polar ions. The frequency window of $\ 0.3\div 3\,\mathrm{THz}$ is less
explored, hence the quest posed by the needs of the emergent THz
technologies in the compact and tunable devices working in this THz
frequency range~\cite{THzBook2012,THzBook2013}.

In this Letter we demonstrate that oscillations of the periodic $180^{\circ
} $ polarization domains in ferroelectric thin films exhibit resonance
behavior in the exactly this $0.3\div 3\,\mathrm{THz}$ frequency range.
Thus, ferroelectric films containing polarization domains are a fertile
ground for designing sub- and low THz radiation range devices.

The regular polarization domain structures, that form in order to cancel a
macroscopic depolarizing field, which would have possessed large energy
proportional to the volume of the system, were first predicted in earlier
works by Landau~\cite{Landau1935,Landau8} and Kittel \cite{Kittel1946} in
the context of ferromagnetic systems. However their existence in the
ferroelectrics have long been considered as barely possible until recent
direct experimental evidences for equilibrium $180^{\circ }$ stripe domains
in strained ferroelectric thin films of PbTiO$_{3}$ (PTO) deposited on the
SrTiO$_{3}$ (STO) substrate \cite{Streiffer2002,Hruszkewycz2013} and in
PTO/STO superlattices~\cite{Zubko2010,Zubko2012} became available. The newly
discovered domain structures exhibited behaviors in a good accord with the
theoretical predictions~\cite%
{Bratkovsky2000,Kornev2004,Lukyanchuk2005,DeGuerville2005,Aguado-Fuente2008,
Lukyanchuk2009} and were considered to be extremely suitable for the future
nanodomain-based electronics \cite{Catalan2012}.

\begin{figure}[b]
\includegraphics [width=2.4cm] {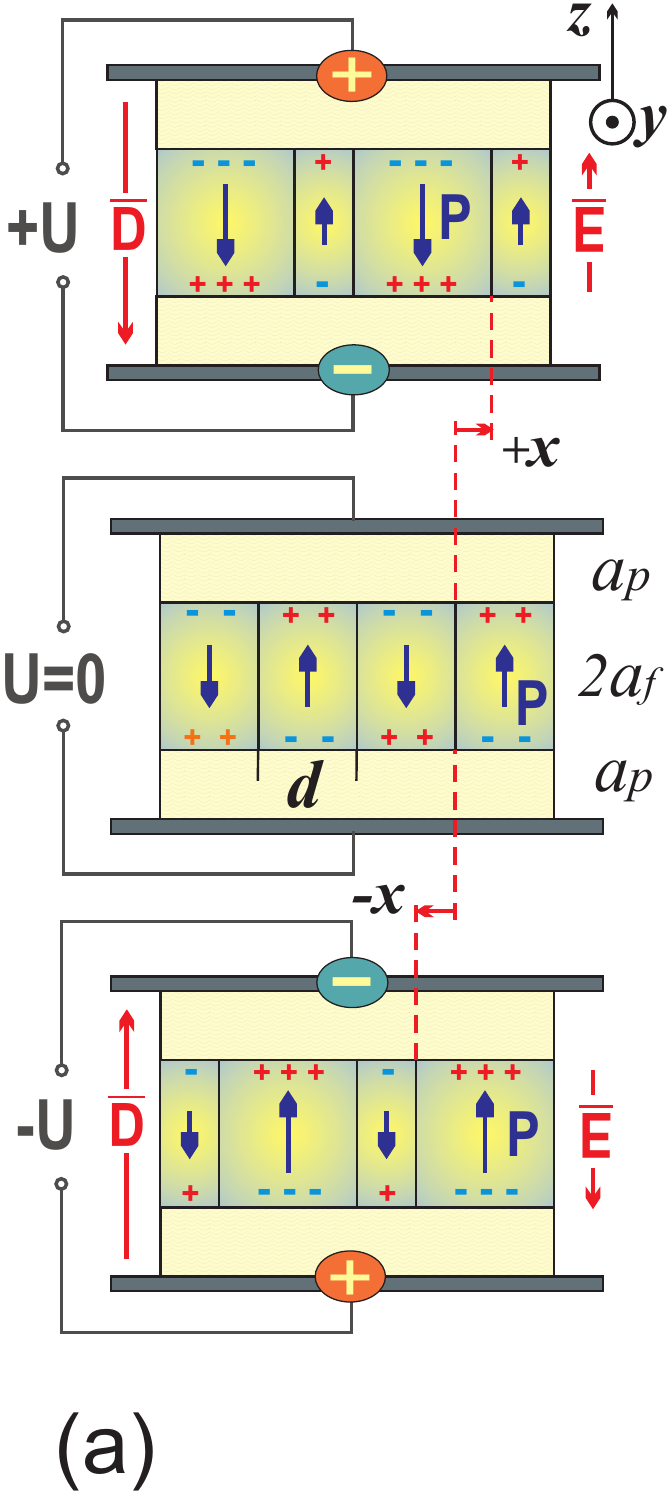} \hspace{0.5cm} %
\includegraphics [width=5.5cm] {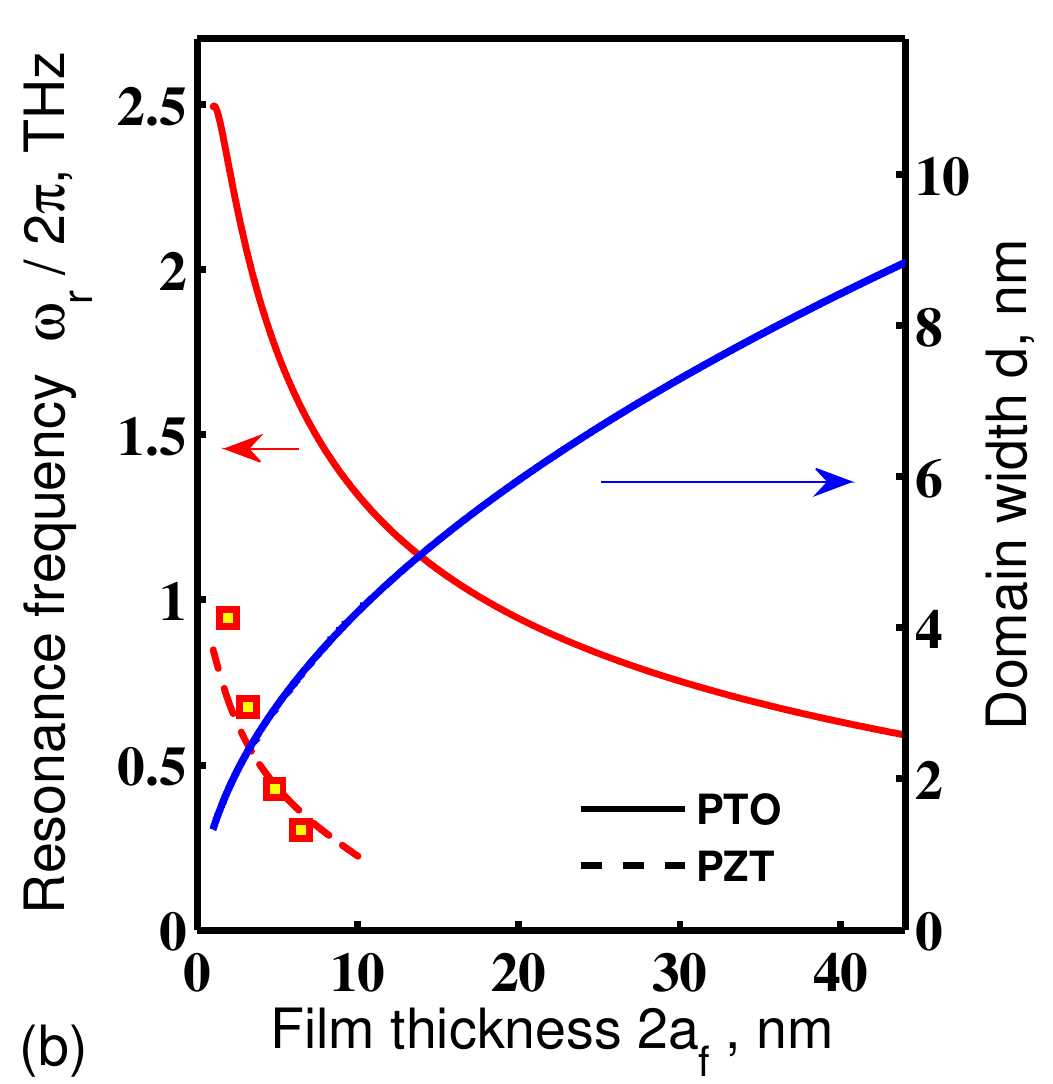}
\caption{(a)\,Oscillation of the periodic domain structure in the sandwiched
ferroelectric layer under the applied ac field. (b)\,Dependence of the
domain resonance frequency, $\protect\omega_r$ (for PTO and PZT) and of the
domain width, $d$, (for PTO) on the film thickness, $2a_f$. Filled squares
show the results of \emph{ab-initio} simulations for PZT \protect\cite%
{Zhang2011}. }
\label{FigOmegaD}
\end{figure}

Although the low-frequency dissipative motion of DWs in macroscopic
ferroelectric samples has been receiving a theoretical attention~\cite%
{Sidorkin}, the crossover from the relaxation dynamics to the resonance was
revealed only recently in the very thin ($2\div 7\,\mathrm{nm}$) films of
PbTi$_{0.6}$Zr$_{0.4}$O$_{3}$ (PZT) by \textit{ab-initio} simulations \cite%
{Zhang2011} at sub-THz frequencies of $0.3\div 1\,\mathrm{THz}$. In what
follows we develop the theory of collective vibrations of periodic $%
180^{\circ }$ domain structure and calculate its dynamical permittivity.
This allows us to generalize the results of~\cite{Zhang2011} for the
arbitrary ferroelectric films to optimize the oscillation parameters for the
best application conditions. In particular, (see Fig.~\ref{FigOmegaD}b), we
demonstrate that the domain resonance in traditional PTO/STO systems can be
achieved at higher frequencies, with smaller relative damping and for wider
range of film thickness. We propose that these oscillations can be detected
by methods of reflection-absorbtion spectroscopy~\cite{Tolstoy2003}.

A graphic illustration of the response mechanism of a domain structure to
the applied ac electric field is presented in Fig.~\ref{FigOmegaD}a. A
ferroelectric film of the thickness $2a_{f} $ is sandwiched between two
paraelectric layers of the thickness $a_{p}$ and placed into the biased
capacitor. Then a periodic structure of the domain stripes aligned along the 
$\mathbf{y}$-axis forms in the ferroelectric layer. The polarization axis $%
\mathbf{z}$ is perpendicular to the film plane, while the in-plane DW motion
occurs along the $\mathbf{x}$-axis~\cite{Brierley2014}. Note that the
described geometry is formally equivalent to the repeating system of
paraelectric and ferroelectric layers of thickness of $2a_{f}$ and $2a_{p}$ 
\cite{Lukyanchuk2005}. Therefore, our consideration is equally applicable to
the corresponding ferroelectric/paraelectric superlattices.

The dielectric properties of the system are described by the intrinsic
permittivities of the ferroelectric film along and across the polar axis, $%
\varepsilon _{\parallel }$, $\varepsilon _{\perp }$, and by the permittivity
of the paraelectric layers, $\varepsilon _{p}$, that are assumed to be
almost frequency independent in the interesting for us sub- and low-THz
range. In the absence of the field, the up- and down- oriented spontaneous
polarizations of alternating domains, $\pm P_{s}$, compensate each other.
The equilibrium domain width, $d$ is given by the famous Landau-Kittel
square root low \cite{Landau1935,Kittel1946,Landau8}, presented in the
universal form \cite{DeGuerville2005}, 

\begin{equation}
d\simeq \left( \frac{\varepsilon _{\perp }}{\varepsilon _{\Vert }}\right)
^{1/4}\sqrt{3.53\varsigma \,\delta \,\left( 2a_{\mathrm{f}}\right) }
\label{width}
\end{equation}%
and exemplified for PTO films in Fig.~\ref{FigOmegaD}b. Here the DW
thickness, $\delta $, is about 1\thinspace nm~\cite{Meyer2002}, $\varsigma
=1+\varepsilon _{\mathrm{p}}/\left( \varepsilon _{\Vert }\varepsilon _{\perp
}\right) ^{1/2}$ and the values of intrinsic permittivities along and across
the polarization direction, $\varepsilon _{\Vert }$, $\varepsilon _{\perp }$
and of paraelectric layer, $\varepsilon _{\mathrm{p}}$, are specified below;
in the experimental range $1\leqslant \varsigma \leqslant 4$.

The applied field induces the net polarization of the ferroelectric layer, $%
\overline{\mathbf{P}}\parallel \mathbf{z}$, that can be conveniently
decomposed into two contributions, $\overline{\mathbf{P}}=\overline{\mathbf{P%
}}_{i}+\overline{\mathbf{P}}_{dw}$. The intrinsic contribution, $\overline{%
\mathbf{P}}_{i}$ is provided by increasing of the up- and decreasing of the
down-oriented polarization inside the alternating domains with the
conservation of the domain width, $d$. Another contribution, $\overline{%
\mathbf{P}}_{dw}$, is due the DWs motion with the extension of the up- and
contraction of the down- oriented domains, while the spontaneous
polarization inside domains preserves. As a driving parameter, it is
convenient to use the $x$-$y$ plane coarse-grained electrostatic induction, $%
\overline{\mathbf{D}}\parallel \mathbf{z}$, which conserves through the
paraelectric-ferroelectric interface and therefore is constant across the
whole system. Calculation of the linear response of the DW structure, $%
\overline{\mathbf{P}}_{dw}=\gamma \overline{\mathbf{D}}$ is one of our tasks
and will be done below.

The resonance of the system is expected when the frequency of the driving ac
field matches the frequency of proper oscillations of DWs, which are
described within the harmonic oscillator approximation \cite{Sidorkin}:\ \ 
\begin{equation}
\mu \overset{..}{x}(t)+\eta \overset{.}{x}(t)+k{x}(t)=2P_{s}E_{D}(t),
\label{Oscil}
\end{equation}%
where $x$ is the coordinate of alternating DW displacements (Fig.~\ref%
{FigOmegaD}a), coefficients $k$, $\mu $ and $\eta $ are calculated per unit
of DW area, and $2P_{s}E_{D}(t)$ is the pressure of the induction-induced
driving field $\overline{\mathbf{E}}_{D}=\left( \varepsilon _{0}\varepsilon
_{\Vert }\right) ^{-1}\overline{\mathbf{D}}$, forcing DW to move to flip the
surrounding polarization from $-P_{s}$ to $+P_{s}$ \cite{DWmotion}. \textit{%
Ab initio} simulations indeed reveal the oscillation dynamics of DWs in the
sub-THz range and allow for evaluation of the coefficients of Eq.~(\ref%
{Oscil}) for thin films of PZT \cite{Zhang2011}. \ In what follows we will
discuss the physical origin of these parameters and generalize the results
of \cite{Zhang2011} for arbitrary ferroelectric material.

The stiffness constant $k$ is of a purely electrostatic origin and,
therefore, is not too sensitive to the profile of DW. Stiffness arises from
the restoring force, trying to diminish the depolarizing field, $E_{dep}$
penetrating the film. The latter is caused by depolarizing surface
extra-charge $\sigma =\pm \Delta P_{s}$, induced by the domain spontaneous
polarization excess $\Delta P_{s}=2\frac{x}{d}P_{s}$ and is directed
oppositely to $\Delta P_{s}$: $E_{dep}=-\sigma /\varepsilon _{0}=-\frac{x}{%
d\varepsilon _{0}}2P_{s}$. The depolarizing field and the intrinsic field
together form the total field, $E=E_{D}+E_{dep}$, inside the ferroelectric
slab. We evaluate the depolarization energy as the electrostatic energy of
the parallel-plate capacitor with the plate area $S$, carrying the
uniformly-distributed charge $Q=\sigma S$, and having the capacity $%
C=\varepsilon _{0}\varepsilon _{\Vert }S/(2a_{f})$ as: 
\begin{equation}
W=\frac{Q^{2}}{2C}=\frac{1}{2\varepsilon _{0}\varepsilon _{\Vert }}%
(2a_{f})\left( 2\frac{x}{d}P_{s}\right) ^{2}S\,.
\end{equation}%
The corresponding energy per unit area of the displaced DW is $w=\frac{d}{%
S(2a_{f})}W$. Relating it with the harmonic oscillator stiffness energy $%
\frac{1}{2}kx^{2}$ we express the coefficient $k$ as: 
\begin{equation}
k=\frac{4P_{s}^{2}}{d\varepsilon _{0}\varepsilon _{\Vert }}g(z),
\end{equation}%
where the correction factor $g(z)$ 
\begin{equation}
g(z)=\frac{1}{z}\ln \cosh z
\end{equation}%
with 
\begin{equation}
z=z=({\pi }\varsigma /{2})\left( \varepsilon _{\perp }/\varepsilon
_{\parallel }\right) ^{1/2}(2a_{\mathrm{f}}/d)
\end{equation}%
was introduced to account for the non-uniform stepwise distribution of
depolarization surface charges at termination of alternating domains \cite%
{Sidorkin}.

The other two coefficients of Eq.~(\ref{Oscil}), effective DW mass, $\mu $,
and the viscosity, $\eta $, are related with the motion of the
material-constituent ions during the variation of the dynamic polarization.
Kittel evaluated the DW mass, considering the flipping of polar ions located
across propagating DW~\cite{Kittel1951}. Subsequent calculations, however,
demonstrated that the effective mass is much larger than that obtained there
and depends on the film thickness~\cite{Sidorkin,Zhang2011}. This is
explained by the larger fraction of participating ions~\cite{Zhang2011}
located in the "soften" polarization profile~\cite%
{Lukyanchuk2005,Lukyanchuk2009} across the entire domain and by the
piezoelectric effect of the depolarization field~\cite{Sidorkin}. We adopt
here the calculated in~\cite{Zhang2011} effective mass, approximately fitted
as $\mu \,\mathrm{[kg/m}^{2}\mathrm{]}\simeq 1.3\sqrt{2a_{f}\,\mathrm{[nm]}}%
\times 10^{-9}$. Another parameter, viscosity, $\eta $, is expressed via DW
relaxation time, $\tau \simeq \mu /\eta $, which was also calculated in~\cite%
{Zhang2011} for PZT films. It was shown, in particular, that $\tau $ is
naturally related with the soft-mode relaxation of polar ions, $\tau _{i}$,
that is only few times shorter then $\tau $~\cite{Pinning}.

Equation~(\ref{Oscil}) establishes the response of the DW vibration, $%
x(t)=x_{\omega }e^{-i\omega t}$, to the periodically applied field, $\mathbf{%
E}_{D}(t)=\mathbf{E}_{D\,\omega }e^{-i\omega t}$. This allows for finding
the dynamical relation between Fourier components of the driving induction $%
\overline{\mathbf{D}}_{\omega }=\varepsilon _{0}\varepsilon _{\Vert }\mathbf{%
E}_{D\,\omega }$ and DW contribution to the ferroelectric layer
polarization, $\overline{\mathbf{P}}_{dw\,\omega }$, that is nothing but the
described above polarization excess, $\Delta P_{\omega }\,\mathbf{=}2P_{s}%
\frac{x_{\omega }}{d}$. We obtain the characteristic Lorentzian resonance
profile: 
\begin{equation}
\overline{\mathbf{P}}_{dw\,\omega }=\gamma (\omega )\,\overline{\mathbf{D}}%
_{\omega },  \label{PgD}
\end{equation}%
with 
\begin{equation}
\gamma (\omega )=\frac{g^{-1}\,\omega _{0}^{2}}{\omega _{0}^{2}-\omega
^{2}-i\Gamma \omega },  \label{gomega}
\end{equation}%
where 
\begin{equation}
\omega _{0}=\sqrt{\frac{k}{\mu }}=\left( \frac{4P_{\mathrm{s}}^{2}g}{\mu
d\varepsilon _{0}\varepsilon _{\Vert }}\right) ^{1/2}  \label{gomega0}
\end{equation}%
is the natural oscillator frequency. Parameters $g$ and $d$ depend on $2a_{%
\mathrm{f}}$. The amplitude of the response function, $|\gamma (\omega )|$,
reaches its maximum at the resonance frequency $\omega _{\mathrm{r}%
}^{2}=\omega _{0}^{2}-\Gamma ^{2}/2$, see (see Fig.\thinspace 3a).

Equation (\ref{gomega0}) enables optimization of the materials parameters
and the film thickness to make sure that $\omega _{\mathrm{r}}$ falls within
the desired THz frequency range. In particular, for the strained films of
PTO with the high spontaneous polarization, $P_{s}\simeq 0.65\,\mathrm{C\,m}%
^{-2}$, see Ref.\,[27], 
relatively low permittivities $\varepsilon _{\Vert }\simeq 100$, $%
\varepsilon _{\perp }\simeq 30$, soft mode damping factor $\Gamma \simeq 20\,%
\mathrm{cm}^{-1}$ ($0.6\,\mathrm{THz}$)\thinspace \cite{Hlinka2011} and $\mu 
$ defined as above, the resonance frequency $\omega _{r}$ decreases and
spans the range from $1.5$ to $0.75\,\mathrm{THz}$ when $2a_{\mathrm{f}}$
increases from $10$ to $40\,\mathrm{nm}$. 

We find the damping frequency, $\omega _{\mathrm{d}}^{2}=\omega
_{0}^{2}-\Gamma ^{2}/4$, of the attenuated oscillations of domains in PTO, $%
x(t)=x_{0}e^{-(\Gamma /2)\,t}\sin \omega _{\mathrm{d}}t$, which is slightly
larger than $\omega _{\mathrm{d}}$, see Fig.\thinspace 3b. Remarkably, our
formulas perfectly describe the results of \textit{ab-initio} simulations of
DWs oscillations in PZT ultrathin films\cite{Zhang2011}. The calculated
damping frequency, $\omega _{\mathrm{d}}$ is shown by the dashed line in the
Fig.\thinspace 3b, the symbols display the results of simulations. Here we
used the following parameters for PZT: $P_{\mathrm{s}}\simeq 0.40\,\mathrm{%
C\,m}^{-2}$ Ref.\thinspace \lbrack 27], 
$\varepsilon _{\Vert },\varepsilon _{\perp }\simeq 350$ , $\Gamma \simeq 27%
\mathrm{cm}^{-1}$ ($0.8\,\mathrm{THz}$)\thinspace \cite{Buixaderas2010} and
the same $\mu $. At the same time, the resonance frequency of PZT, $\omega _{%
\mathrm{r}}$, drops rapidly with the thickness and vanishes above
4\thinspace nm.

\begin{figure}[h]
\centering
\includegraphics [width=4cm] {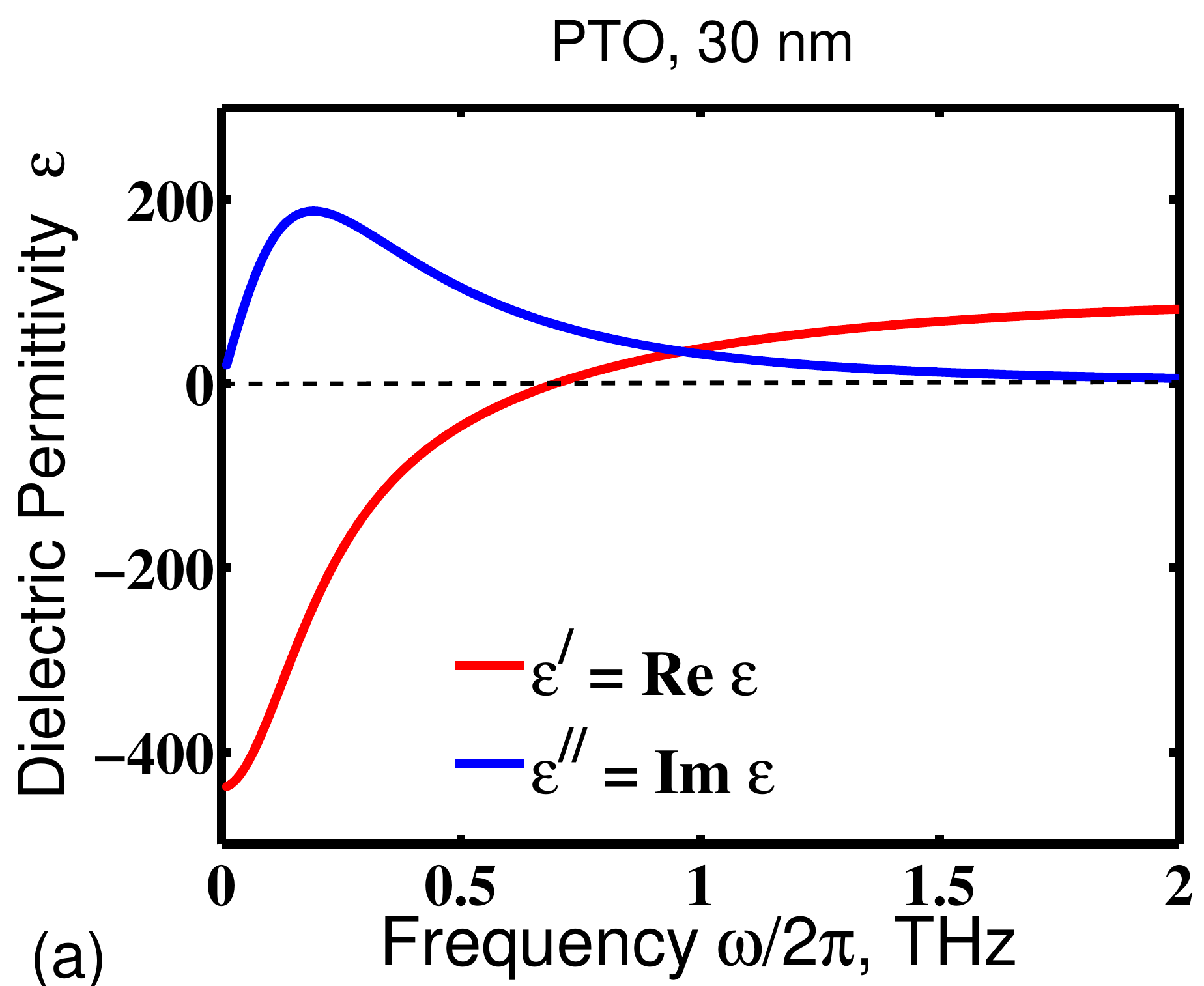} \centering
\includegraphics [width=4cm] {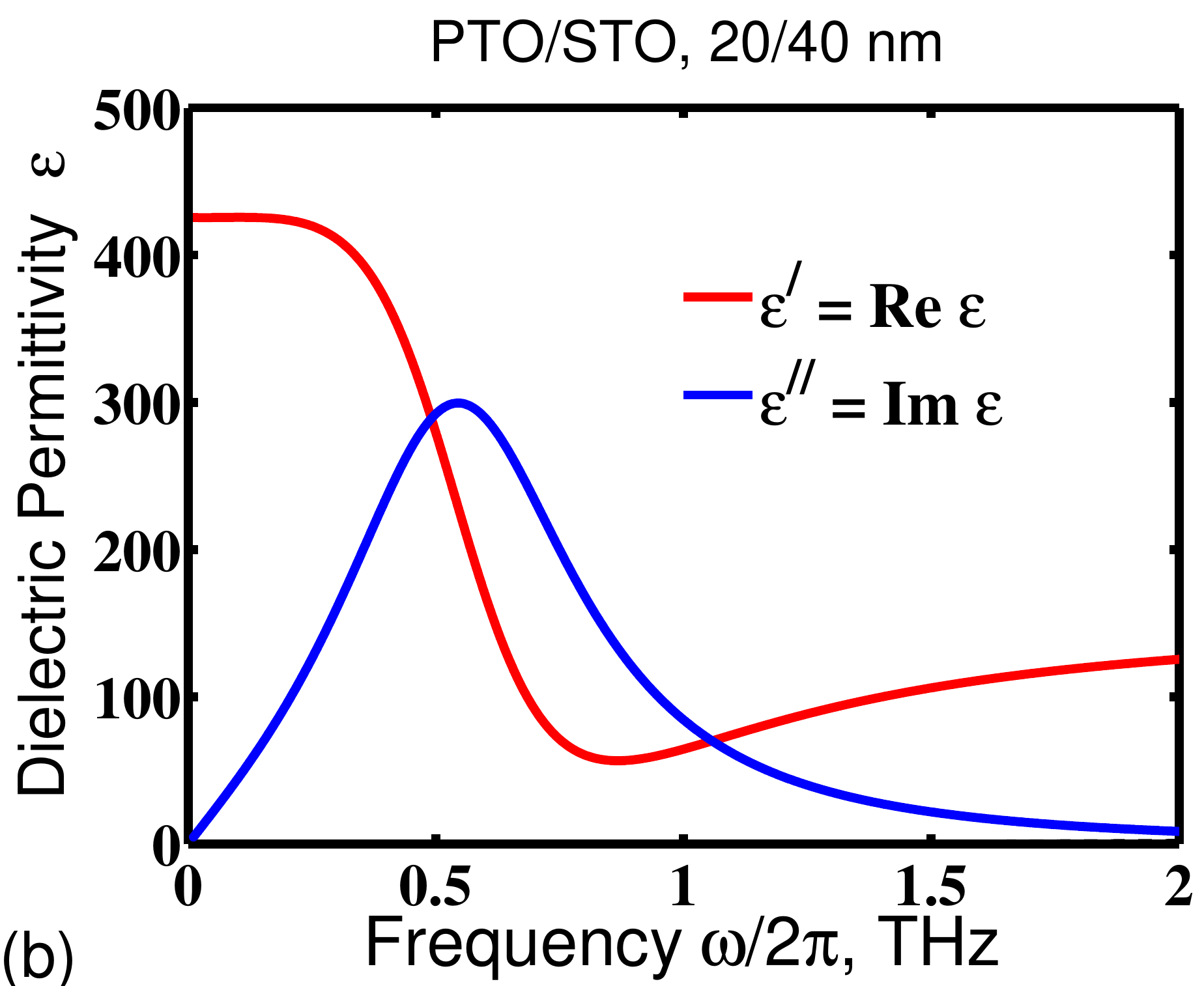} \centering
\caption{(a) Dynamic dielectric permittivity of the $30\,\mathrm{nm}$ thick
film of PTO. (b) The same for either the $20/40\,\mathrm{nm}$ sandwich of
PTO/STO layers or for the equivalent superlattice.}
\label{FigEps}
\end{figure}

Having established the dynamical response of the DW-induced polarizations on
the applied induction (\ref{PgD}), we find the effective dielectric
permittivity, $\varepsilon _{f}(\omega )$. The latter is defined as
coefficient of proportionality between induction and the average electric
field \textit{inside} the layer, $\overline{\mathbf{D}}_{\omega
}=\varepsilon _{0}\varepsilon _{f}(\omega )\overline{\mathbf{E}}_{\omega }$.
Taking into account that $\overline{\mathbf{D}}_{\omega }=\varepsilon
_{0}\varepsilon _{\Vert }\overline{\mathbf{E}}_{\omega }+\overline{\mathbf{P}%
}_{dw\,\omega }$, where the first term is due to intrinsic contribution, $%
\overline{\mathbf{P}}_{i}$ we obtain \cite{Kopal1999}: 
\begin{equation}
\varepsilon _{f}(\omega )=\frac{\varepsilon _{\Vert }}{1-\gamma (\omega )}.
\label{epsilonf}
\end{equation}

The frequency dependence of $\varepsilon _{f}(\omega )$ for $30$ $\mathrm{nm}
$ film of PTO (Fig.~\ref{FigEps}a) exhibits striking frequency dependence:
its real part is negative at low frequencies, and then becomes positive in
sub-THz region. This peculiarity already noted in \cite{Bratkovsky2006}\ for
the static $\varepsilon _{f}(0)$, is explained as a result of the opposite
orientation of depolarization field $\overline{\mathbf{E}}_{dep}$ with
respect to $\overline{\mathbf{P}}\parallel $ $\overline{\mathbf{D}}$. This
field appears to be stronger then the \textquotedblleft correctly" oriented
driving field $\overline{\mathbf{E}}_{D}=\left( \varepsilon _{0}\varepsilon
_{\Vert }\right) ^{-1}\overline{\mathbf{D}}$ and results in orientation of
the total field inside ferroelectric slub, $\overline{\mathbf{E}}=\overline{%
\mathbf{E}}_{D}+\overline{\mathbf{E}}_{dep}$ against $\overline{\mathbf{D}}$.

This negative-$\varepsilon $ phenomenon does not lead  however to
thermodynamic contradiction, since the described domain structure can be
observed only when the thickness of the paraelectric buffer, $a_{p}$, is
larger then the domain width, $d$, and domain depolarization stray fields do
not interact with electrodes \cite{Mokry2004}. Then, only the effective
permittivity, $\overline{\varepsilon }$, defined through the total capacity
of the system $C=\varepsilon _{0}\overline{\varepsilon }\frac{S}{%
2a_{f}+2a_{d}}$ makes sense. Under this condition, $\overline{\varepsilon }%
(\omega )$ can be decomposed into two in-series contributions from para- and
ferroelectric layers:%
\begin{equation}
\frac{1}{\overline{\varepsilon }(\omega )}=\frac{\alpha _{p}}{\varepsilon
_{p}}+\frac{\alpha _{f}}{\varepsilon _{f}(\omega )},\qquad \alpha _{f,p}=%
\frac{a_{f,p}}{a_{f}+a_{p}}.  \label{epstot}
\end{equation}%
Plot of the $\overline{\varepsilon }(\omega )$ dependence for PTO/STO $30/50$
$\mathrm{nm}$ system with $\varepsilon _{p}=\varepsilon _{\text{STO}}\left(
1\,\mathrm{THz}\right) \simeq 200+30i$ at $300\,\mathrm{K}$ \cite%
{Ostapchuk2002} (Fig.~\ref{FigEps}b) indeed demonstrates the positive $%
\mathrm{Re\,}\overline{\varepsilon }$ at low $\omega $ with the peak at $%
\simeq 0.5\,\mathrm{THz}$. Making use of expressions (\ref{gomega})-(\ref%
{epstot}) we find: 
\begin{equation}
\overline{\varepsilon }(\omega )=\overline{\varepsilon }(\infty )+\frac{%
\Delta \overline{\varepsilon }\,\omega _{0}^{\prime 2}}{\omega _{0}^{\prime
2}-\omega ^{2}-i\Gamma \omega }=\frac{\omega _{0}^{2}-\omega ^{2}-i\Gamma
\omega }{\omega _{0}^{\prime 2}-\omega ^{2}-i\Gamma \omega }\overline{%
\varepsilon }(\infty )\,,
\end{equation}%
with $\Delta \overline{\varepsilon }=\overline{\varepsilon }(0)-\overline{%
\varepsilon }(\infty )$, and 
\begin{gather}
\frac{1}{\overline{\varepsilon }(0)}=\frac{\alpha _{p}}{\varepsilon _{p}}+%
\frac{\alpha _{f}}{\varepsilon _{\Vert }}\left( 1-\gamma _{0}\right) , \\
\frac{1}{\overline{\varepsilon }(\infty )}=\frac{\alpha _{p}}{\varepsilon
_{p}}+\frac{\alpha _{f}}{\varepsilon _{\Vert }}.  \notag
\end{gather}%
The effective oscillator frequency: $\omega _{0}^{\prime }<\omega _{0}$
satisfies the Lyddane-Sachs-Teller relation 
\begin{equation}
\frac{\omega _{0}^{\prime 2}}{\omega _{0}^{2}}=\frac{\overline{\varepsilon }%
(\infty )}{\overline{\varepsilon }(0)}=1-\alpha _{f}\frac{\varepsilon
(\infty )}{\varepsilon _{\Vert }}\gamma _{0}  \label{LST}
\end{equation}%
and can be tuned by the geometrical parameters $\alpha _{p}$, $\alpha _{f}$.

\begin{figure}[h]
\centering
\includegraphics [width=8cm] {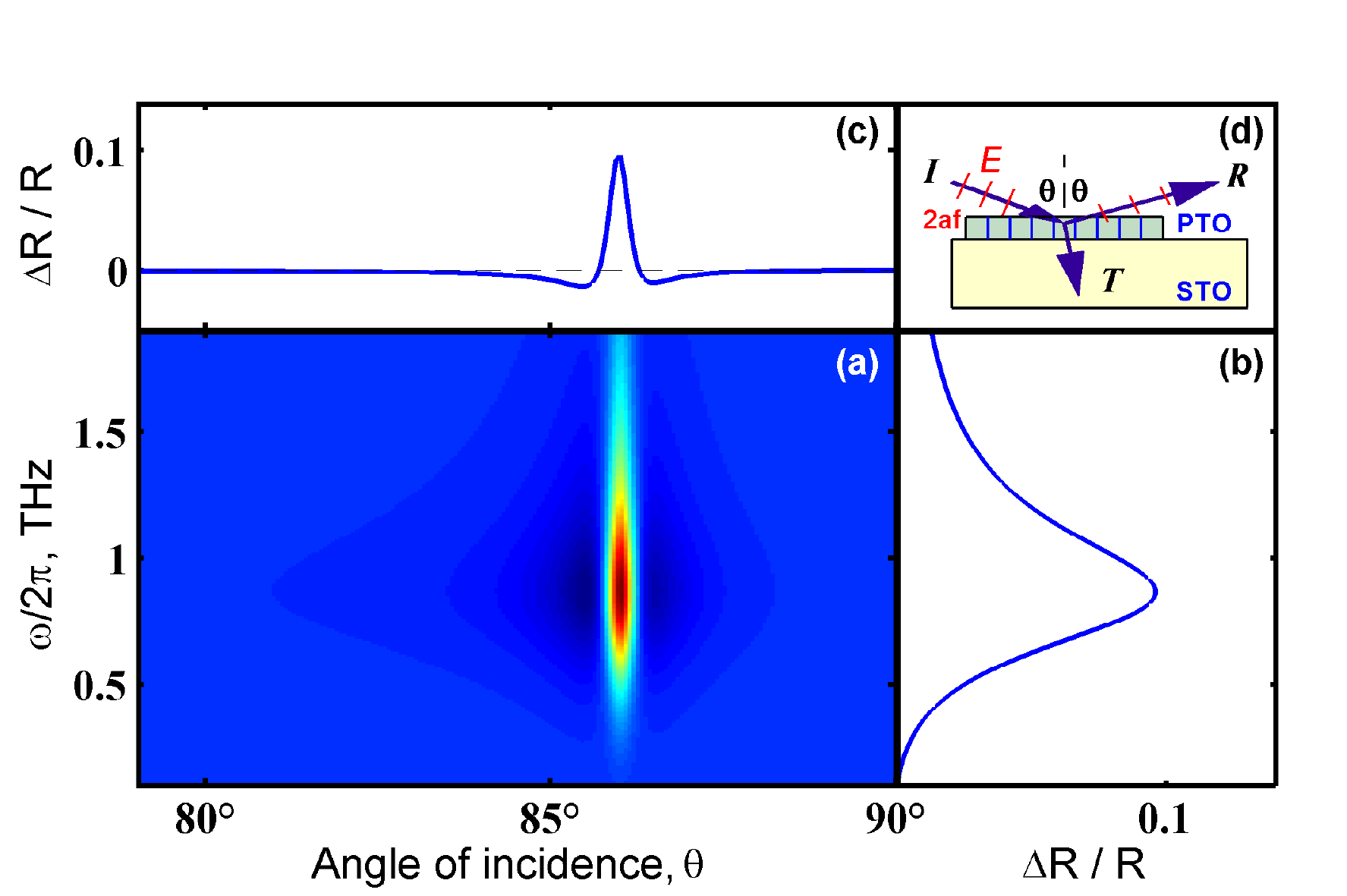}
\caption{Calculated reflectivity, $\Delta R / R$, for the $p$-polarized beam
of sub-THz radiation of $2a_f\simeq30\,\mathrm{nm}$ PTO film with domains,
deposited on the thick STO substrate, as function of the incidence angle, $%
\protect\theta$, and the beam frequency, $\protect\omega$. The resonance
enhancement is observed at $\protect\omega_r/2\protect\pi \simeq 0.75\,%
\mathrm{THz}$ and at the Brewster incidence angle $\protect\theta _{B}\simeq
86^{\circ}$. (a) Two-dimensional color map of $\Delta R / R$ dependence of $%
\protect\theta$ and $\protect\omega$. (b) Cross-cut of $\Delta R / R$ at $%
\protect\theta=\protect\theta _{B}$. (c) Cross-cut of $\Delta R / R$ at $%
\protect\omega=\protect\omega_{r}$. (d) The geometry of the experiment. }
\label{FigDRall}
\end{figure}

The classical impedance-spectroscopy setup of Fig.~\ref{FigOmegaD}a, used so
far for calculations of $\varepsilon _{f}(\omega )$, is not suitable for the
experimental study in sub- and low-THz region, where the far-infrared and
THz-optics can be explored \cite{Tolstoy2003,THzBook2012,THzBook2013}. The
low-THz resonance in ultrathin films of $2\div 10\,\mathrm{nm}$ , where the
resonance frequency is only slightly smaller then the frequency of the soft
mode $E(TO_{1})$, can be also captured by Raman spectroscopy. In the
reflection-absorption geometry shown in Fig.~\ref{FigDRall}, the incident $p$%
- (in-plane) polarized THz beam is reflected from the STO substrate through
the PTO film and the intensity of the outcoming radiation is measured. The
dynamics of domain structure can be detected by its interaction with the $%
\mathbf{z}$-component of the field, $\mathbf{E}$, when the reflected beam
passes twice through the PTO layer.

The THz reflectance of the thin film with $2a_{f}<\lambda =2\pi c/\omega $,
where the wavelength $\lambda \simeq 1\div 0.15\,\mathrm{mm}$ for $0.3\div
2\,\mathrm{THz}$ radiation, deposited on the thick substrate with $%
a_{p}>\lambda $, is calculated from $\varepsilon _{f}(\omega )$ as \cite%
{Tolstoy2003}:

\begin{equation}
R^{(p)}\approx R_{0}^{(p)}\left\vert 1-\frac{2a_{f}}{\lambda }\frac{8\pi
\,\cos \theta \sin ^{2}\theta }{\left( \cos \theta -\varepsilon
_{p}^{-1/2}\sin \theta \right) ^{2}}\mathrm{Im}\frac{1}{\varepsilon
_{f}(\omega )}\right\vert ,
\end{equation}%
where $\theta $ is the incidence angle and 
\begin{equation}
R_{0}^{(p)}=\left\vert \frac{\varepsilon _{p}\cos \theta -\sqrt{\varepsilon
_{p}-\sin ^{2}\theta }}{\varepsilon _{p}\cos \theta +\sqrt{\varepsilon
_{p}-\sin ^{2}\theta }}\right\vert ^{2}
\end{equation}%
is the reflectance of the substrate alone.

To identify the contribution of the domain structure we consider the thin $%
30\;\mathrm{nm}$ film of PTO deposited on thick, $\sim 0.5\,\mathrm{mm}$
substrate of STO. The frequency and the incidence angle dependencies of
reflectivity, $\Delta R/R=(R_{0}^{(p)}-R^{(p)})/R_{0}^{(p)}$, \cite%
{Tolstoy2003} is shown in Fig.~\ref{FigDRall}. The domain-provided
enhancement of reflectivity of about $\Delta R/R\simeq 0.1$, is observed at $%
\omega /2\pi \simeq 0.75\,\mathrm{THz}$ at the Brewster incidence angle $%
\theta _{B}\approx \arctan \left( \mathrm{Re\,}\varepsilon _{p}\right)
^{1/2}\simeq 86^{\circ }$, when the reflectance of $p$-wave from the
substrate alone, $R_{0}^{(p)}$, \ is as small as the imaginary part of $%
\varepsilon _{\text{STO}}$. The incident beam almost tangential to the
surface can be realized by placing the high-refractive Otto prism at the top
of the sample \cite{Tolstoy2003}. Importantly, even better result can be
achieved by using the PTO/STO superlattice instead of PTO\ monolayer. The
reflectivity, $\Delta R/R$, will increase proportionally to the number of
PTO layers and can become giant at $\theta =\theta _{B}$.

To conclude, we calculated the dynamic dielectric permittivity $\varepsilon
_{f}(\omega )$ of the regular structure of $180^{\circ }$ domains in
strained ferroelectric films and superlattices. We demonstrated that the
collective vibrational mode with the resonance frequency $\omega _{r}/2\pi
\simeq 0.3\div 3\,\mathrm{THz}$ can be detected in PTO/STO systems by means
of the reflection-absorption spectroscopy. This unique property makes
ferroelectric films a promising candidate for compact and tunable devices
working in the sub- and low $\mathrm{THz}$ range.

\acknowledgments

We are delighted to thank A.\thinspace Razumnaya for providing experimental
parameters. This work was supported by IRSES-SIMTECH and ITN-NOTEDEV FP7
mobility programs and by the U.S. Department of Energy, Office of Science,
Materials Sciences and Engineering Division.

\subsection{Supplementary Material: \newline
Static Permittivity of Ferroelectric Film with $180^\circ$ Domains}

Static permittivity of the ferroelectric film with domains, sandwiched
between two paraelectric layers was calculated by A.~Kopal, P.~Mokr\'{y},
J.~Fousek and T.~Bahn\'{\i}k \cite{Kopal1999} 
as: 
\begin{equation}
\varepsilon _{\mathrm{eff}}=\varepsilon _{\mathrm{z}}^{(2)}\frac{D}{%
d^{\prime }}\left( 1+B\frac{\varepsilon _{\mathrm{z}}^{(2)}}{\varepsilon _{%
\mathrm{z}}^{(1)}}\right) ^{-1}+\left[ \left( 1+B\frac{\varepsilon _{\mathrm{%
z}}^{(2)}}{\varepsilon _{\mathrm{z}}^{(1)}}\right) \left[ \frac{B}{\left(
1+B\right) \varepsilon _{\mathrm{z}}^{(1)}}-\frac{2\ln 2}{R_{\mathrm{eq}%
}^{0}\left( 1+B\right) \left( g^{(1)}+g^{(2)}\right) }\left( 1+B\frac{%
\varepsilon _{\mathrm{z}}^{(2)}}{\varepsilon _{\mathrm{z}}^{(1)}}\right) %
\right] \right] ^{-1},
\end{equation}%
In our notations,%
\begin{gather}
\varepsilon _{\mathrm{eff}}=\varepsilon _{\mathrm{tot}}(0),\quad \varepsilon
_{\mathrm{z}}^{(1)}=\varepsilon _{\mathrm{p}},\quad \varepsilon _{\mathrm{z}%
}^{(2)}=\varepsilon _{\parallel },\quad g^{(1)}+g^{(2)}=\varepsilon _{%
\mathrm{p}}+\left( \varepsilon _{\parallel }\varepsilon _{\perp }\right)
^{1/2}=\varsigma \left( \varepsilon _{\parallel }\varepsilon _{\perp
}\right) ^{1/2}, \\
\frac{d^{\prime }}{D}=\frac{2a_{\mathrm{f}}}{2a_{\mathrm{p}}+2a_{\mathrm{f}}}%
=\alpha _{\mathrm{f}},\quad B=\frac{2a_{\mathrm{p}}}{2a_{\mathrm{f}}}=\frac{%
\alpha _{\mathrm{p}}}{\alpha _{\mathrm{f}}},\quad R_{\mathrm{eq}}^{0}=\frac{{%
\pi }}{{2}}\frac{2a_{\mathrm{f}}}{d},  \notag
\end{gather}%
Eq.~(\ref{EpsTotKopal}) can be written as: 
\begin{equation}
\varepsilon _{\mathrm{tot}}(0)=\left( 1+\frac{2a_{\mathrm{p}}}{2a_{\mathrm{f}%
}}\frac{\varepsilon _{\parallel }}{\varepsilon _{\mathrm{p}}}\right)
^{-1}\left( \frac{\varepsilon _{\parallel }}{\alpha _{f}}+\left[ \frac{%
\alpha _{\mathrm{p}}}{\varepsilon _{\mathrm{p}}}-\frac{\alpha _{\mathrm{f}}}{%
\varepsilon _{\parallel }}\frac{4\ln 2}{{\pi }\varsigma }\left( \frac{%
\varepsilon _{\parallel }}{\varepsilon _{\perp }}\right) ^{1/2}\frac{d}{2a_{%
\mathrm{f}}}\left( 1+\frac{2a_{\mathrm{p}}}{2a_{\mathrm{f}}}\frac{%
\varepsilon _{\parallel }}{\varepsilon _{\mathrm{p}}}\right) \right]
^{-1}\right) .
\end{equation}

The inverse permittivity, after some algebra, can be decomposed onto two
in-series contributions from paraelectric and ferroelectric layers, 
\begin{equation}
\frac{1}{\varepsilon _{\mathrm{tot}}(0)}=\frac{\alpha _{\mathrm{p}}}{%
\varepsilon _{\mathrm{p}}}+\frac{\alpha _{\mathrm{f}}}{\varepsilon _{\mathrm{%
f}}(0)}
\end{equation}%
where the permittivity of ferroelectric layer, 
\begin{equation}
\varepsilon _{\mathrm{f}}(0)=\varepsilon _{\parallel }-\frac{{\pi }\varsigma 
}{4\ln 2}\left( \frac{\varepsilon _{\perp }}{\varepsilon _{\parallel }}%
\right) ^{1/2}\frac{2a_{\mathrm{f}}}{d}\varepsilon _{\parallel },
\end{equation}%
has the positive (intrinsic) contribution and the negative (DW-provided)
contribution.


\begin{thebibliography}{99}
\bibitem{Pinning} In general, pinning forces are relavant for dynamics of DW
at frequences comparable to or below the pinning frequences and neglibible
at higher frequences. According to \cite{Sidorkin}, dynamical interaction of
DWs with impurities are important in the GHz region and, therefore, can be
neglected in THz nanodynamics.
\bibitem{Eps} The data are given for ceramic samples. The values for
strained films can be slightly different.
\end{thebibliography}
\end{document}